\documentclass[fullpaper,twocolumn]{jpsj2}
\usepackage{txfonts}
\usepackage{bm}

\title{Theory of Orbital Susceptibility in the Tight-Binding Model: \\ Corrections to the Peierls Phase  
}

\author{
Hiroyasu Matsuura\thanks{matsuura@hosi.phys.s.u-tokyo.ac.jp} and Masao Ogata
}
\inst{
$^1$ Department of Physics, University of Tokyo, Hongo, Bunkyo-ku, Tokyo 113-0033, Japan \\
} 

\abst{
An extended formula for orbital susceptibility including corrections of the Peierls phase is introduced.
By using the new developed formula, the orbital susceptibility of benzene is estimated analytically on the basis of the $\pi$ electron approximation.
As a result, it is found that the orbital susceptibility is 1.2 times larger than that estimated only from the Peierls phase.
The Coulomb interaction dependence of the orbital susceptibility of benzene is also discussed by using exact diagonalization.   
It is found that the absolute value of the orbital susceptibility decreases as the Coulomb interaction increases, while the ratio of the orbital susceptibility with and without the corrections of the Peierls phase increases.
Finally, we discuss the orbital susceptibility of a single-band tight-binding model on a square lattice.
We clarify that the correction of the Peierls phase is comparable to the Landau--Peierls orbital susceptibility and that it corresponds to the Fermi sea term.
}

\begin{document}
\maketitle
\section{Introduction}
Orbital magnetism derived from the motion of an electron in a magnetic field has attracted interest from the time of the development of quantum mechanics.
However, owing to the complex matrix elements between atomic orbitals or Bloch bands, it is difficult to understand it exactly.

Historically, after the orbital magnetism was discussed in the case of an isolated atom by applying quantum mechanics, the orbital magnetism of molecules and crystals was discussed.
However, because the amplitude of the vector potential depends on the distance from the origin, it was difficult to estimate the orbital magnetism in large molecules and crystals.
Peierls introduced a way to avoid the problem, the so-called Peierls phase, and derived the Landau--Peierls formula, $\chi_{{\rm LP}}$, for a single-band model, which is used to estimate the orbital magnetism of a crystal~\cite{Peierls}. 
In the same period, London also introduced a similar method and calculated the orbital susceptibility of molecules on the basis of the $\pi$-electron approximation~\cite{London}.   
After that, Pople derived a general formulation to estimate the orbital magnetism of complex molecules on the basis of the Peierls phase~\cite{Pople}.

In the case of crystals, there have been several discussions on the derivation of orbital susceptibility~\cite{Luttinger,Wilson,Adams,Kjeldaas,Kohn,HS2,HS3,Roth,Blount,Wannier,Ichimaru,HFKubo,HFKubo2,Fukuyama}. 
Finally, Fukuyama derived a simple general formula to calculate the orbital susceptibility in crystals in terms of Green's function without using the Peierls phase~\cite{Fukuyama}.
Recently, by applying Fukuyama's formula to Bloch bands, a general formula in terms of Bloch wave functions has been introduced~\cite{Ogata}.    
This formula shows that there are several contributions to orbital susceptibility in addition to $\chi_{{\rm LP}}$~\cite{HS2,HS3,Ogata}.
Actually, for single-band models, it was shown that the additional contributions to the orbital susceptibility are comparable to $\chi_{{\rm LP}}$~\cite{Ogata2}.
This is rather surprising since it has been long believed that $\chi_{{\rm LP}}$ obtained from the Peierls phase~\cite{Raoux} is predominant, at least, in the single-band tight-binding model.
This means that the Peierls phase is an approximation and the corrections to the Peierls phase play important roles.
Actually in Peierls's original paper, an approximation is used for simplicity to introduce the Peierls phase:
integration with respect to ${\bf r}$ is performed, then, the mean value of two sites ${\bf r} =({\bf R}_i+{\bf R}_j)/2$ is used to avoid complicated integrals.

In this paper, we develop a formula that is exact up to the first order with respect to the overlap integral between neighboring atomic orbitals by extending Pople's formalism.
In this exact treatment, we show that there are several correction terms in addition to the Peierls phase.
Then, we apply the obtained exact formula to benzene as an example.
Benzene is a simple molecule, and its orbital magnetism has been discussed theoretically since 80 years ago.
For example, the orbital susceptibility has been estimated on the basis of the $\pi$-electron approximation, the Peierls phase, and Coulomb interaction. 
However, it is found that the estimated value is 74 \verb|%| of the experimental value~\cite{Fujii}.
In this paper, we show that the estimated value based on the new formula is 1.2 times larger than the previous result.
In addition, we consider the effect of the Coulomb interaction.

To clarify the meaning of the corrections of the Peierls phase, we apply the formula to a single-band tight-binding model on a square lattice.
The orbital susceptibility of a square lattice was calculated on the basis of the Peierls phase in a previous study, and $\chi_{{\rm LP}}$ was obtained~\cite{Raoux}. 
However, in this paper, we obtain not only $\chi_{{\rm LP}}$ but also the correction terms on the basis of the new formula.
We clarify that the correction corresponds to the Fermi sea term.

This paper is organized as follows.
In Sect.\ 2, to obtain the orbital susceptibility exactly, we derive a general formula for orbital susceptibility by extending Pople's formulation based on the Peierls phase.  
Next, in Sect.\ 3, to demonstrate the corrections of the Peierls phase, we calculate the orbital susceptibility of benzene analytically on the basis of the $\pi$-electron approximation.
The estimated value based on the new formula is 1.2 times larger than the previous result based on the Peierls phase.
We also discuss the orbital susceptibility of benzene by considering the Coulomb interaction.
Finally, in Sect.\ 4, we estimate the orbital susceptibility of a single-band tight-binding model on a square lattice.
It is found that the corrections of the Peierls phase give contributions to the orbital susceptibility comparable to $\chi_{{\rm LP}}$.
We also clarify that the correction of the Peierls phase corresponds to the Fermi sea term in the tight-binding model.

\section{General Formulation of Orbital Susceptibility in the Tight-Binding Model}
By extending Pople's formulation~\cite{Pople}, we derive a general formula to estimate the orbital susceptibility.
The Hamiltonian describing the motion of an electron under a magnetic field is    
\begin{eqnarray}
 \hat{H}({\bf r}) = \frac{1}{2m}\bigr({\bf p} -\frac{e}{c}{\bf A} \bigr)^2 + \sum_{j}V({\bf r}_{j}) , \label{H1} 
\end{eqnarray}
where ${\bf A}$ is the magnetic vector potential, ${\bf r}$ is the position of an electron with charge $e<0$, ${\bf r}_{j} \equiv {\bf r} -{\bf R}_j$, ${\bf R}_j$ represents the position of a nucleus with effective nuclear charge $Z^{*}e$, and  $V({\bf r}_j)$ is the Coulomb potential from the nuclear charge given as 
\begin{eqnarray}
V({\bf r}_{j}) = -\frac{Z^{*}e^2}{r_j},
\end{eqnarray}
with $r_j = |{\bf r}-{\bf R}_j|$.
In this letter, we use a symmetric gauge given as
\begin{eqnarray}
{\bf A} = \frac{1}{2}{\bf H}\times {\bf r}, \hspace{0.5cm}{\bf H}= (0,0,H). \label{sym_gauge}
\end{eqnarray}
Peierls introduced a wave function at ${\bf r}_i = {\bf r}-{\bf R}_i$ as\begin{eqnarray}
\psi_{m}({\bf r}_i) = \exp{\bigr( \frac{ie}{c\hbar}{\bf A}_i \cdot {\bf r} \bigr)}\phi_{m}({\bf r}_i),
\end{eqnarray}
where ${\bf A}_i$ is the vector potential at ${\bf R}_{i}$ defined as ${\bf A}_i = \frac{1}{2}{\bf H}\times {\bf R}_{i}$ and $\phi_{m}({\bf r}_i)$ is the $m$th atomic wave function satisfying
\begin{eqnarray}
\bigr[ \frac{{\bf p}^2}{2m} + V({\bf r}_i) \bigr] \phi_{m}({\bf r}_i) =  \epsilon_m \phi_m({\bf r}_i),
\end{eqnarray}
where $\epsilon_m$ is the $m$th eigenvalue.

Generally speaking, an orthogonalized wave function based on atomic orbitals under a magnetic field is given by a linear combination of $\psi_{m}({\bf r}_i)$ as follows~\cite{Lowdin}:
\begin{eqnarray}
\Phi_{m}({\bf r}_i) = \sum_{l,k}C_{mi,lk}\psi_{l}({\bf r}_k),  \label{wave1}
\end{eqnarray}
where $C_{mi,lk}$ is a coefficient, and this function satisfies
\begin{eqnarray}
\int d{\bf r} \Phi_{m}^{*}({\bf r}_i)\Phi_{n}({\bf r}_j) = \delta_{mn}\delta_{ij}. \label{orth1}
\end{eqnarray}
As discussed in Ref. \citen{Lowdin}, we introduce thegoverlap integralhbetween $\psi_{m}({\bf r}_i)$ and $\psi_{n}({\bf r}_j)$ as follows:
\begin{eqnarray}
S_{mi,nj} = \int d{\bf r} \psi_{m}^{*}({\bf r}_i)\psi_{n}({\bf r}_j) -\delta_{mn}\delta_{ij}. \label{over1}
\end{eqnarray}
Using eqs.\ (\ref{orth1}) and (\ref{over1}), 
we can rewrite eq.\ (\ref{wave1}) as
\begin{eqnarray}
\Phi_{m}({\bf r}_i) &=& \sum_{l,k}(\delta_{ml}\delta_{ik} + S_{lk,mi})^{-\frac{1}{2}}\psi_{l}({\bf r}_k). 
\end{eqnarray}
When the overlap integral is smaller than unity, we expand $\Phi_m({\bf r}_i)$ in terms of the overlap integral as 
\begin{eqnarray}
\Phi_{m}({\bf r}_i) &=& \psi_{m}({\bf r}_i) -\frac{1}{2}\sum_{l,k}S_{lk,mi}\psi_{l}({\bf r}_k) + \cdots. \label{expand1}
\end{eqnarray}

In the following, we calculate Hamiltonian matrix elements in terms of the orthogonalized wave functions $\Phi_m$(${\bf r}_i$).
This gives a tight-binding model.
Then, by diagonalizing the Hamiltonian, we calculate the ground-state energy as a function of the magnetic field and obtain the susceptibility.
Using eq.\ (\ref{expand1}), the matrix element of $\hat{H}({\bf r})$ is obtained as
\begin{eqnarray}
&&\int d{\bf r} \Phi_{m}^*({\bf r}_i) \hat{H}({\bf r}) \Phi_{n}({\bf r}_j) = \int d{\bf r} \psi_{m}^*({\bf r}_i) \hat{H}({\bf r}) \psi_{n}({\bf r}_j) \nonumber \\
&& -\frac{1}{2}\sum_{l,k}\biggr[ S_{lk,mi}^{*} \int d{\bf r} \psi_{l}^*({\bf r}_k) \hat{H}({\bf r}) \psi_{n}({\bf r}_j) \nonumber \\
&& + S_{lk,nj} \int d{\bf r} \psi_{m}^*({\bf r}_i) \hat{H}({\bf r}) \psi_{l}({\bf r}_k) \biggr] + \cdots  .  \label{eq2}
\end{eqnarray}
When ${\bf R}_i = {\bf R}_j$, this matrix element gives the on-site energy,
while it gives the hopping integral in the tight-binding model when ${\bf R}_i \neq {\bf R}_j$.
The first term in eq.\ (\ref{eq2}) is transformed as
\begin{eqnarray}
&& \int d{\bf r} \psi_{m}^*({\bf r}_i) \hat{H}({\bf r}) \psi_{n}({\bf r}_j) \nonumber \\
&=& \int d{\bf r} e^{-\frac{ie}{c\hbar}({\bf A}_{i} - {\bf A}_{j} )\cdot {\bf r} }\phi_{m}^*({\bf r}_i)\tilde{H}({\bf r})\phi_{n}({\bf r}_j), \\
&=& e^{-i\Phi_{ij}}\int d{\bf r}e^{i\chi_{ij}({\bf r})}\phi_{m}^*({\bf r}_i) \tilde{H}({\bf r})\phi_{n}({\bf r}_j), \label{pierls}
\end{eqnarray}
where 
\begin{eqnarray}
\tilde{H}({\bf r}) \equiv  \frac{1}{2m}\biggr({\bf p} -\frac{e}{c}({\bf A}-{\bf A}_{j}) \biggr)^2 + \sum_{j}V({\bf r}_j). \label{newH}
\end{eqnarray}
The phases $\Phi_{ij}$ and $\chi_{ij}({\bf r})$ are defined as
\begin{eqnarray}
\Phi_{ij} &=&  \frac{e}{c\hbar}({\bf A}_i - {\bf A}_{j}) \cdot (\frac{{\bf R}_{i}+{\bf R}_{j}}{2}) \nonumber\\
          &=& \frac{h}{a^2}\bigr[{\bf R}_{i}\times {\bf R}_{j}\bigr]_z, \label{Peierls} \\
\chi_{ij}({\bf r}) &=&  -\frac{e}{c\hbar}({\bf A}_i - {\bf A}_{j}) \cdot ({\bf r}-\frac{{\bf R}_{i}+{\bf R}_{j}}{2}) \nonumber \\
&=& - \frac{h}{2a^2}\bigr[({\bf R}_{i} -{\bf R}_{j})\times ({\bf r}_i+{\bf r}_j)\bigr]_z,  \end{eqnarray}
where $\bigr[ \bf X \bigr]_z$ represents the $z$-component of vector $\bf X$ and $h$ is the dimensionless parameter
\begin{eqnarray}
h = \frac{eH}{2c\hbar}a^2,
\end{eqnarray}
with $a$ being the nearest-neighbor distance, $a \equiv |{\bf R}_{i}-{\bf R}_j|$.
Here, we have considered only the nearest-neighbor sites ${\bf R}_{i}$ and ${\bf R}_{j}$.
However, the extension to longer-range terms is straightforward.
In Pople's paper~\cite{Pople}, $\chi_{ij}({\bf r})$ was neglected by assuming $\bf r$ to be the mean value for two sites, i.e., ${\bf r}=({\bf R}_{i}+{\bf R}_{j})/2$. 
In contrast, we calculate the orbital susceptibility exactly taking account of $\chi_{ij}({\bf r})$ in this paper. 

Generally speaking, the Peierls phase is defined as
\begin{eqnarray}
\exp{\biggr( \frac{ie}{c\hbar}\int_{{\bf R}_j}^{{\bf R}_i} {\bf A} \cdot d{\bf l} \biggr) } \label{PP}
\end{eqnarray}
for ${\bf R}_{i} \neq {\bf R}_{j}$.
The line integral in eq.\ (\ref{PP}) depends on the path, but it is conventionally calculated by assuming a straight path as follows:
\begin{eqnarray}
&&\frac{ie}{c\hbar}\int_{{\bf R}_j}^{{\bf R}_i} {\bf A} \cdot d{\bf l} \nonumber \\ 
&=& \frac{ie}{c\hbar}\int_{0}^{1}\bigr[ {\bf A}_j + s({\bf A}_i -{\bf A}_j)\bigr] \cdot ({\bf R}_i -{\bf R}_j)ds, \\
&=& \frac{ie}{c\hbar}\frac{({\bf A}_i +{\bf A}_j)}{2} \cdot ({\bf R}_i -{\bf R}_j), \\
&=& -\frac{ie}{c\hbar}({\bf A}_i -{\bf A}_j) \cdot \frac{({\bf R}_j +{\bf R}_i)}{2},
\end{eqnarray}
where we have used the symmetric gauge in eq.\ (\ref{sym_gauge}).
Thus, $\exp{(-i\Phi_{ij})}$ in eq.\ (\ref{pierls}) is simply the Peierls phase.

The second and third terms in eq.\ (\ref{eq2}) are transformed in the same way.
As a result, eq.\ (\ref{eq2}) is written as
\begin{eqnarray}
{\rm eq.\ (\ref{eq2})} &=& e^{-i\Phi_{ij}}\langle e^{i\chi_{ij}({\bf r})}\tilde{H}({\bf r}) \rangle_{mi,nj} \nonumber \\ 
&& -\frac{1}{2}\sum_{l,k} \bigr[ S_{lk,mi}^{*}e^{-i\Phi_{kj}}\langle e^{i\chi_{kj}({\bf r})}\tilde{H}({\bf r}) \rangle_{lk,nj} \nonumber \\
&& + S_{lk,nj}e^{-i\Phi_{ik}}\langle e^{i\chi_{ik}({\bf r})} \tilde{H}({\bf r})  \rangle_{mi,lk}\bigr], \label{eq1} 
\end{eqnarray}
and the overlap integral defined in eq.\ (\ref{over1}) is
\begin{eqnarray}
S_{lk,nj} = e^{-i\Phi_{kj}}\langle e^{i\chi_{kj}({\bf r})} \rangle_{lk,nj} - \delta_{ln}\delta_{kj},
\end{eqnarray}
where $\langle \hat{\it O} \rangle_{mi,nj}$ is defined as 
\begin{eqnarray}
\langle \hat{O} \rangle_{mi,nj} = \int d{\bf r} \phi_{m}^*({\bf r}_i) \hat{\it O} \phi_{n}({\bf r}_j).
\end{eqnarray}

It is important to remark here that the vector potential ${\bf A}$ appears in eq.\ (\ref{eq1}) only in the form of ${\bf A} -{\bf A}_j$ or ${\bf A}_i -{\bf A}_j$.
Since ${\bf A} -{\bf A}_j$ and ${\bf A}_i -{\bf A}_j$ are rewritten in terms of relative coordinates, the problems of calculating orbital susceptibility due to the real-space coordinate ${\bf r}$ are removed.

First, let us calculate the site-diagonal term, i.e., for the case with ${\bf R}_i = {\bf R}_j$.
In this case, the second and third terms in eq.\ (\ref{eq1}) become second order with respect to the overlap integral.
Therefore, they can be neglected when we consider contributions up to the first order.
Since $\Phi_{ii}= \chi_{ii}({\bf r})=0$, we simply have eq.\ (\ref{eq1}) $= \langle \tilde{H}({\bf r})\rangle_{mi,ni}$.
When $\tilde{H}({\bf r})$ is divided into three terms as 
\begin{eqnarray}
 \tilde{H}({\bf r}) &=&  {H}_{0}({\bf r}) + {H}_{1}({\bf r}) + {H}_{2}({\bf r}), \\
{H}_{0}({\bf r}) &=& \frac{{\bf p}^2}{2m}+ \sum_{j}V({\bf r}_j), \\
{H}_{1}({\bf r}) &=&  -\frac{e}{mc} ({\bf A}-{\bf A}_j) \cdot {\bf p} 
  = -\frac{eH}{2mc}\bigr[ {\bf r}_j \times {\bf p}\bigr]_z  \\
{H}_{2}({\bf r}) &=&  \frac{e^2}{2mc^2}({\bf A}-{\bf A}_j)^2, 
=\frac{e^2H}{8mc^2}(x_j^2+y_j^2),
\end{eqnarray}
where $\bigr[ {\bf r}_i \times{\bf p} \bigr]_z$ is the $z$-component of ${\bf r}_i \times {\bf p}$ and $x_j$ and $y_j$ are the $x$- and $y$-components of ${\bf r}_j$, respectively, we obtain
\begin{eqnarray}
\langle \tilde{H}({\bf r}) \rangle_{mi,ni} &=& \langle H_0({\bf r}) \rangle_{mi,ni} +\langle H_1({\bf r}) \rangle_{mi,ni}+\langle H_2({\bf r}) \rangle_{mi,ni}, \\
&=& \epsilon_{m}\delta_{m,n} + \int \phi_{m}^{*}({\bf r}_i)\sum_{j \neq i}V({\bf r}_j)\phi_{n}({\bf r}_i) d{\bf r} \nonumber \\
 &&-\frac{eH}{2mc}\langle L_z \rangle_{mi,ni} +\frac{e^2H^2}{8mc^2}\langle x_i^2 +y_i^2 \rangle_{mi,ni}. \nonumber \\
 \label{diagonal}
\end{eqnarray}
The third term leads to the Van Vleck susceptibility in the atomic limit and the fourth term gives the atomic diamagnetism.
Note that even in the presence of overlap integrals between the neighboring atomic orbitals, the contributions of the Van Vleck susceptibility and atomic diamagnetism do not change up to the first order with respect to the overlap integral~\cite{Ogata,Ogata2}.

For the matrix elements with ${\bf R}_i \neq {\bf R}_j$, the general formula is somewhat complicated.
Since $\chi_{ij}({\bf r})$ is proportional to $h$, we expand $e^{i\chi_{ij}({\bf r})}$ in terms of $h$ up to the second order.
Furthermore, $\langle e^{i\chi_{ij}({\bf r})}\tilde{H}({\bf r}) \rangle_{lk,nj}$ in the second term of eq.\ (\ref{eq1}) should be evaluated as the site diagonal $\langle \tilde{H}({\bf r}) \rangle_{lj,nj}$ since we consider terms up to the first order with respect to the overlap integral.
As a result, we obtain  
\begin{eqnarray}
\rm{eq.\ (\ref{eq1})} &=&  e^{-i\Phi_{ij}}\biggr[
\langle e^{i\chi_{ij}({\bf r})}\tilde{H}({\bf r}) \rangle_{mi,nj} \nonumber \\
 &&-\frac{1}{2}\sum_{l} \bigr[ \langle e^{i\chi_{ij}({\bf r})}\rangle_{lj,mi}^{*} \langle \tilde{H}({\bf r})  \rangle_{lj,nj} \nonumber \\
&& +\langle e^{i\chi_{ij}({\bf r})}\rangle_{li,nj} \langle \tilde{H}({\bf r}) \rangle_{mi,li}\bigr] 
\biggr], \\
&=& e^{-i\Phi_{ij}}\bigr[ t_{mi,nj} + t_{mi,nj}^{\prime}h + t_{mi,nj}^{\prime\prime}h^2 +o(h^3)\bigr], \label{hop1} 
\end{eqnarray}
where $t_{mi,nj}$, $t_{mi,nj}^{\prime}h$, and $t_{mi,nj}^{\prime\prime}h^2$ are given by
\begin{eqnarray}
t_{mi,nj} &=& \langle H_0({\bf r}) \rangle_{mi,nj} \nonumber \\
&& -\frac{1}{2}\sum_{l} \bigr[\langle 1 \rangle_{lj,mi}^{*} \langle H_0({\bf r})  \rangle_{lj,nj} + \langle 1 \rangle_{li,nj} \langle H_0({\bf r}) \rangle_{mi,li} \bigr], \label{t0}\\
t_{mi,nj}^{\prime}h &=& \langle H_1({\bf r}) \rangle_{mi,nj} \nonumber \\
&& -\frac{1}{2}\sum_{l} \bigr[\langle 1 \rangle_{lj,mi}^{*} \langle H_1({\bf r})  \rangle_{lj,nj} + \langle 1 \rangle_{li,nj} \langle H_1({\bf r}) \rangle_{mi,li}  \bigr] \nonumber \\
&& +\langle i\chi_{ij}({\bf r})H_0({\bf r}) \rangle_{mi,nj} \nonumber \\
&& -\frac{1}{2}\sum_{l} \bigr[ \langle i\chi_{ij}({\bf r})\rangle_{lj,mi}^{*} \langle H_0({\bf r})  \rangle_{lj,nj} + \langle i\chi_{ij}({\bf r})\rangle_{li,nj} \langle H_0({\bf r}) \rangle_{mi,li} \bigr], \nonumber \\
&& \label{t1} \\
t_{mi,nj}^{\prime\prime}h^2 &=& \langle H_2({\bf r}) \rangle_{mi,nj} \nonumber \\
&& -\frac{1}{2}\sum_{l} \bigr[\langle 1 \rangle_{lj,mi}^{*} \langle H_2({\bf r})  \rangle_{lj,nj} + \langle 1 \rangle_{li,nj} \langle H_2({\bf r}) \rangle_{mi,li}  \bigr] \nonumber \\
&& +\langle i\chi_{ij}({\bf r})H_1({\bf r}) \rangle_{mi,nj} \nonumber \\
&& -\frac{1}{2}\sum_{l} \bigr[ \langle i\chi_{ij}({\bf r})\rangle_{lj,mi}^{*} \langle H_1({\bf r})  \rangle_{lj,nj} +\langle i\chi_{ij}({\bf r})\rangle_{li,nj} \langle H_1({\bf r}) \rangle_{mi,li} \bigr] \nonumber \\
&& -\frac{1}{2}\biggr[ \langle \chi_{ij}^2({\bf r})H_0({\bf r}) \rangle_{mi,nj} \nonumber \\
&& -\frac{1}{2}\sum_{l} \bigr[ \langle \chi_{ij}^2({\bf r})\rangle_{lj,mi}^{*} \langle H_0({\bf r})  \rangle_{lj,nj} + \langle \chi_{ij}^2({\bf r})\rangle_{li,nj} \langle H_0({\bf r}) \rangle_{mi,li} \bigr] \biggr]. \nonumber \\
&& \label{t2}
\end{eqnarray}
Equation (\ref{hop1}) is the general formula of the hopping integrals in the tight-binding model with a magnetic field.
The first term in eq.\ (\ref{hop1}) is the conventional hopping term with the Peierls phase.
The remaining two terms are correction terms to the Peierls argument, which represent the modifications of the hopping integrals due to the magnetic field.

By diagonalizing the eigenvalue equation under the magnetic field, the total energy is obtained as
\begin{eqnarray}
E_{\rm tot}(H) = \sum_{j}E_{j}(H).
\end{eqnarray}
where $E_{j}(H)$ is the $j$th eigenvalue and $\sum_{j}$ is the summation up to the highest occupied orbital.
In the following, we calculate the total energy $E_{\rm tot}(H)$ up to the second order with respect to the magnetic field $H$ (or $h$).
The orbital susceptibility $\chi$ is obtained as
\begin{eqnarray}
\chi = -\frac{\partial^2 E_{{\rm tot}}(H)}{\partial H^2}\biggr|_{H=0}. 
\end{eqnarray}

\section{Orbital Susceptibility of Benzene}
As a simple example, we calculate the orbital susceptibility of the $\pi$-electron in benzene on the basis of the above formalism. 
In benzene, six carbon atoms are located at ${\bf R}_1=(a,0)$, ${\bf R}_2=(\frac{a}{2},\frac{\sqrt{3}a}{2})$, ${\bf R}_3=(-\frac{a}{2},\frac{\sqrt{3}a}{2})$, ${\bf R}_4=(-{a},0)$, ${\bf R}_5=(-\frac{a}{2},-\frac{\sqrt{3}a}{2})$, and ${\bf R}_6=(\frac{a}{2},-\frac{\sqrt{3}a}{2})$ with $a$ being the distance between two nearest-neighbor carbons.
The $i$th 2p$_\pi$ orbital on a carbon is given by
\begin{eqnarray}
\phi_{2p_\pi}({\bf r}_i) = \bigr(\frac{Z^{*}}{2a_B}\bigr)^{\frac{5}{2}}\frac{z_i}{\sqrt{\pi}}e^{-\frac{Z^*}{2a_B}r_i},
\end{eqnarray}
where $r_i = |{\bf r}-{\bf R}_i|$ and $z_i$ is the $z$-component of ${\bf r}_i$.
We consider only the matrix elements between the nearest neighbor 2p$_{\pi}$ orbitals.
Therefore, the orbital suffix,\ $n$,$m$, is not shown in the following. 

First, since $\hat{L}_z\phi_{2p_\pi}({\bf r}_i)=0$, all the expectation values in eqs.\ (\ref{t0})--(\ref{t2}) involving $H_{1}({\bf r})$ vanish.
Furthermore, $\langle \chi_{ij} \rangle_{ij}=0$ owing to the anisotropy of $\phi_{2p_\pi}({\bf r}_i)$ in the $x$-$y$ plane.
Consequently, eqs.\ (\ref{t0})--(\ref{t2}) are simplified.
In particular, $t_{\pi i,\pi,i+1}^{\prime}$ vanishes, and the transfer integral between ${\bf R}_{i}$ and ${\bf R}_{i+1}$ is obtained as
\begin{eqnarray}
\rm{eq.\ (\ref{eq1})} &=& e^{-i\Phi_{ii+1}}\bigr[ -t  + t_2h^2 \bigr]. 
\end{eqnarray}
Here, $t$ represents the transfer integral independent of the magnetic field, and $t_2$ is the transfer integral induced by the magnetic field.
From eq.\ (\ref{t0}), we obtain 
\begin{eqnarray}
t  &=& -\langle H_{0}({\bf r}) \rangle_{i,i+1}+ S_{i,i+1} \langle H_{0}({\bf r}) \rangle_{i,i}, \label{hop_20} \\
&=& -\langle V({\bf r}_i) \rangle_{i,i+1}+ S_{i,i+1} \langle V({\bf r}_{i+1}) \rangle_{i,i}. \label{hop_2}
\end{eqnarray}
The derivation of eq.\ (\ref{hop_2}) from eq.\ (\ref{hop_20}) is discussed and justified in Ref.\ \citen{Ogata2}.
From a simple calculation, $\langle V({\bf r}_i) \rangle_{i,i+1}$, $\langle V({\bf r}_{i+1}) \rangle_{i,i}$ and $S_{i,i+1}$ are~\cite{Table} 
\begin{eqnarray}
\langle V({\bf r}_i) \rangle_{i,i+1} &=& 
-\frac{(Z^{*})^2e^2}{4a_B}\bigr[ 1+p+\frac{p^2}{3} \bigr]e^{-p}, \\
\langle V({\bf r}_{i+1}) \rangle_{i,i} &=& -\frac{e^2}{2a_B}\frac{Z^*a_B}{a}\biggr[ (2-\frac{3}{p^2}) \nonumber \\
  && +(4+p+\frac{6}{p}+\frac{3}{p^2})e^{-2p} \biggr], \\
S_{i,i+1} &=& \biggr[1+p+\frac{2}{5}p^2 +\frac{1}{15}p^3 \biggr]e^{-p},
\end{eqnarray}
where $p\equiv \frac{Z^{*}a}{2a_B}$.

From eq.\ (\ref{t2}), the transfer integral dependent on the second order of the magnetic field is obtained as
\begin{eqnarray}
 t_2h^2 &=& \langle H_2({\bf r}) \rangle_{i,i+1} -S_{i,i+1}\langle H_2({\bf r}) \rangle_{i,i} \nonumber \\
&& \hspace{-1cm}+ \frac{1}{2}\biggr[ \langle \chi_{ii+1}^2({\bf r}) \rangle_{i,i+1}\langle V({\bf r}_{i+1}) \rangle_{i,i} - \langle \chi_{ii+1}^2({\bf r}) V({\bf r}_i) \rangle_{i,i+1} \biggr]. \label{t2_gene}
\end{eqnarray}
Here, the last term has been derived in a similar way to eq.\ (\ref{hop_2}).
The expectation values are given as 
\begin{eqnarray}
\langle H_2({\bf r}) \rangle_{i,i+1} &=& \frac{e^2H^2}{8mc^2}\langle x_{i}^2 +y_{i}^2 \rangle_{i,i+1}, \\
&=& \frac{e^2}{2a_B}\bigr(\frac{a_B^2 h}{Z^{*}a^2})^2 \biggr[
12+12p+\frac{44}{7}p^2 \nonumber \\
 &&+\frac{16}{7}p^3+\frac{4}{7}p^4+\frac{8}{105}p^5 \biggr]e^{-p}, \\
\langle H_2({\bf r}) \rangle_{i,i} &=& \frac{e^2H^2}{8mc^2}\langle x_{i}^2 +y_{i}^2 \rangle_{i,i} = \frac{e^2}{2a_B}12\bigr(\frac{a_B^2h}{Z^*a^2}\bigr)^2,
\end{eqnarray}
and 
\begin{eqnarray}
\langle \chi_{ii+1}^2({\bf r}) \rangle_{i,i+1} &=& \biggr(\frac{a_Bh}{aZ^*}\biggr)^2\biggr[ 6 + 6p \nonumber \\
 && +\frac{18}{7}p^2 +\frac{4}{7}p^3 +\frac{2}{35}p^4 \biggr]e^{-p}, \\
 \langle \chi_{ii+1}^2({\bf r}) V({\bf r}_i) \rangle_{i,i+1}  
&=& -\frac{e^2}{2a_B}\frac{2a_B^2 h^2}{a^2} \nonumber \\
 && \times \biggr[ 1+p+\frac{2}{5}p^2+\frac{1}{15}p^3 \biggr]e^{-p}. 
\end{eqnarray}

\begin{figure}
\rotatebox{0}{\includegraphics[angle=0,width=1\linewidth]{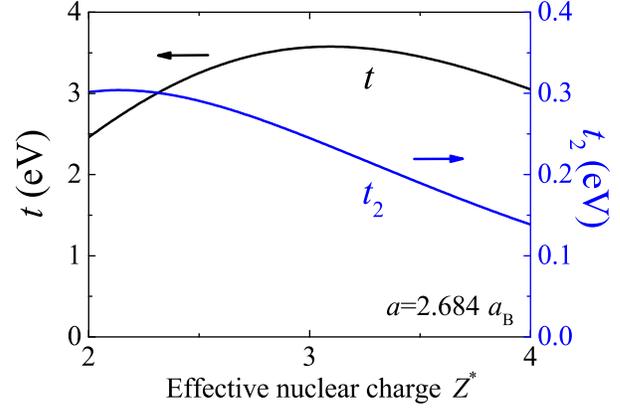}}
\caption{ (Color online) Effective nuclear charge $Z^{*}$ dependences of $t$ (black line) and $t_2$ (blue line) for the distance between carbons $R=2.684a_B$, where $a_B$ is the Bohr radius.
}
\label{hopping}
\end{figure}

Figure \ref{hopping} shows the effective nuclear charge $Z^{*}$ dependences of $t$ and $t_2$ for $a=2.684a_B$, which is the distance between the carbons in benzene.
Since the effective nuclear charge is $Z^{*}=3.25$ by the Slater rule~\cite{Slater}, it is found that the hoppings are $t \simeq 3.6$ eV and $t_2 \simeq 0.22$ eV.

Using the second quantization, the effective Hamiltonian of benzene is expressed as
\begin{eqnarray}
H_{\rm 0} = \sum_{i=1\sim 6,\sigma}\biggr[ e^{-i\Phi_{ii+1}}\bigr[ -t  + t_2h^2 \bigr]c_{i\sigma}^{\dagger}c_{i+1\sigma} +{\rm h.c.} \biggr],
\end{eqnarray}
where $c_{i\sigma}$ ($c_{i\sigma}^{\dagger}$) is an annihilation (creation) operator of the 2p$_{\pi}$ orbital with spin $\sigma$ at the $i$th site.
Here, we neglect the core energy of the 2p$_{\pi}$ orbital because we focus on the orbital magnetism between 2p$_\pi$ orbitals.
 
By diagonalizing this 6 $\times$ 6 matrix,  the six eigenvalues, $\epsilon_1$--$\epsilon_6$, are analytically obtained as
\begin{eqnarray}
\epsilon_1 &=& 2t -(\frac{3}{4}t +2t_2)h^2 +o(h^3), \\
\epsilon_2 &=&  t +\frac{3}{2}th - (\frac{3}{8}t +t_2)h^2+o(h^3), \\
\epsilon_3 &=&  t -\frac{3}{2}th - (\frac{3}{8}t +t_2)h^2,+o(h^3), \\
\epsilon_4 &=&  -t +\frac{3}{2}th + (\frac{3}{8}t +t_2)h^2+o(h^3), \\
\epsilon_5 &=&  -t -\frac{3}{2}th + (\frac{3}{8}t +t_2)h^2+o(h^3), \\
\epsilon_6 &=& -2t +(\frac{3}{4}t +2t_2)h^2+o(h^3). 
\end{eqnarray}
The total energy is given as
\begin{eqnarray}
E = 2\bigr[ -4t + (\frac{3}{2}t +4t_2)h^2 +o(h^3)\bigr],
\end{eqnarray}
where the factor 2 corresponds to the spin degree of freedom.
The orbital susceptibility is given as
\begin{eqnarray}
\chi = -\frac{\partial^2 E}{\partial H^2}\bigr|_{H \rightarrow 0} = -\frac{3e^2a^4}{2c^2\hbar^2}t\bigr( 1 +\frac{8t_2}{3t} \bigr).
\end{eqnarray}
The absolute value of the orbital susceptibility is increased by the effect of $t_2$.
Since $t_2/t \simeq 0.06$ in the realistic region, the orbital susceptibility is 1.2 times larger than that estimated from only the Peierls phase.

Next, we discuss the effect of the Coulomb interaction on the orbital susceptibility of benzene.
The effect of the Coulomb interaction on benzene has previously been discussed in detail~\cite{Matsuura}. 
Here, we use the simple model 
\begin{eqnarray}
H_{\rm tot} = H_{\rm 0} +\sum_{i=1 \sim 6}U n_{i\uparrow}n_{i\downarrow}, 
\end{eqnarray}
where $n_{i\sigma}= c_{i\sigma}^{\dagger}c_{i\sigma}$ and $U$ is the Coulomb interaction of the 2p$_{\pi}$ orbital.

\begin{figure}
\rotatebox{0}{\includegraphics[angle=0,width=1\linewidth]{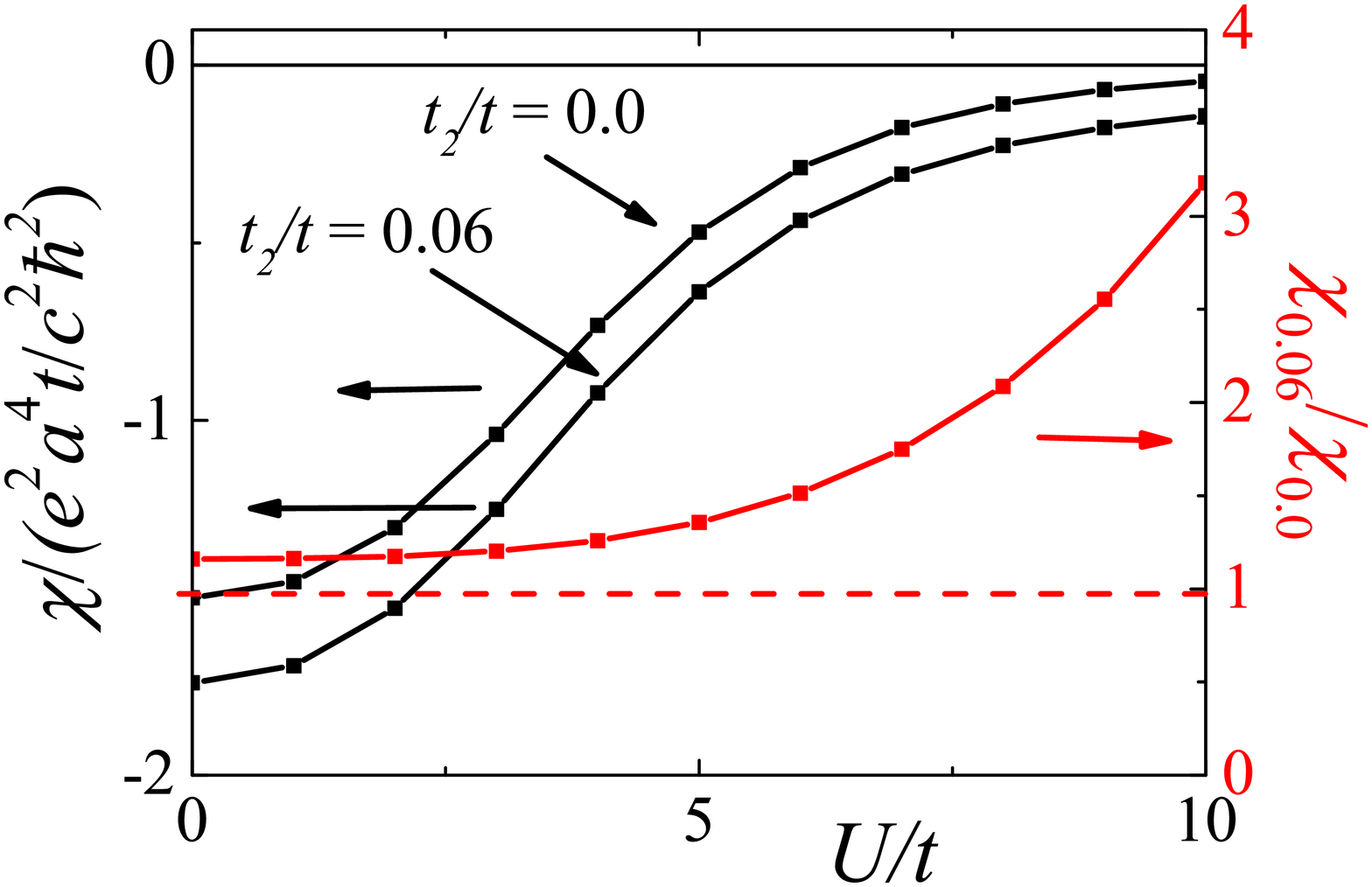}}
\caption{(Color online) Coulomb interaction $U$ dependence of the orbital susceptibility for $t_2/t=0.0$ and $t_2/t= 0.06$.
}
\label{coulomb}
\end{figure}
The orbital susceptibility is obtained by diagonalizing $H_{\rm tot}$ numerically and by estimating the magnetic field dependence of the ground-state energy.
Figure \ref{coulomb} shows the $U$ dependence of the orbital susceptibility for $t_2/t =0$ and $t_2/t = 0.06$.
It is found that as $U$ increases, the absolute value of the orbital susceptibility decreases.
Here the orbital susceptibilities for $t_2/t =0$ and $t_2/t = 0.06$ are denoted as $\chi_{0.0}$ and $\chi_{0.06}$, respectively. 
Then, it is also found that the ratio of the orbital susceptibility of $t_2/t =0.0$ to $t_2/t = 0.06$ increases as $U$ increases (see the red line in Fig. \ref{coulomb} ).
 
As discussed in the introduction, it is known that the orbital susceptibility obtained by the quantum chemical calculation is 74 \verb|%| of the experimental value~\cite{Fujii}.
In the previous calculation, the correction of the Peierls phase ($t_2$) was neglected.
As shown in the above calculations, the orbital susceptibility is increased by the correction.
It is expected that the orbital susceptibility will become closer to the experimental result when the correlation is considered.

\section{Orbital Susceptibility of Square Lattice}
Finally, to clarify the physical meaning of the correction of the Peierls phase, we discuss the orbital susceptibility of a single-orbital tight-binding model on a two-dimensional square lattice.
The Hamiltonian under a magnetic field is 
\begin{eqnarray}
H = \sum_{i,\sigma}e(h)c_{i\sigma}^{\dagger}c_{i\sigma} + \sum_{\langle i,j \rangle, \sigma}\biggr[t(h)e^{-i\Phi_{ij}}c_{i\sigma}^{\dagger}c_{j\sigma} + {\rm h.c.} \biggr],
\end{eqnarray}
where $c_{i\sigma}$ and $c_{i\sigma}^{\dagger}$ are annihilation and creation operators with spin $\sigma$ at $i$th site, respectively.
The first term represents the site diagonal term shown in eq.\ (\ref{diagonal}).
 $\sum_{\langle i,j \rangle}$ in the second term indicates the summation over the nearest-neighbor bonds.
As shown in eq.\ (\ref{diagonal}), $e(h)$ is generally given by
\begin{eqnarray}
e(h) = e_0 +e_1h +e_2h^2.
\end{eqnarray}
For the 2p$_{\pi}$ orbital, $e_1$ vanishes since $\hat{L}_z\phi_{p\pi}=0$, and $e_2$ is given as 
\begin{eqnarray}
e_2 &=& \frac{\hbar^2}{2ma^4}\langle x_i^2 +y_i^2 \rangle_{ii} \\
    &=& \frac{e^2}{2a_B}12\biggr(\frac{a_B^2}{Z^*a^2}\biggr)^2.
\end{eqnarray}
As shown in eq.\ (\ref{hop1}), $t(h)$ is given as
\begin{eqnarray}
t(h) = -t +t_2h^{2}.
\end{eqnarray}

First, the site-diagonal term gives the correction of the ground-state energy as
\begin{eqnarray}
\Delta E_0 = e_2h^2n_e,
\end{eqnarray}
where $n_e$ is the density of electrons.
This gives a contribution to the susceptibility of
\begin{eqnarray}
\chi_0 = -\frac{e^2a^4}{c^2\hbar^2}\frac{12e^2}{2a_B}\bigr(\frac{a_B}{a^2Z^*}\bigr)^2n_e. \label{site_dia}
\end{eqnarray}
This contribution is called the gintrabandhatomic diamagnetism~\cite{Ogata2}, which is naturally connected to the atomic diamagnetism.

We consider the effects of $t$ and $t_2$ separately.
First, $-e^{-i\Phi_{ij}}t$ represents the conventional hopping with the Peierls phase.
This term gives the Landau-Peierls orbital susceptibility as~\cite{Ogata2,Raoux} 
\begin{eqnarray}
\chi_1 = \chi_{LP}
&=& \frac{e^2}{6\hbar^2 c^2}\sum_{{\bf k}}f^{\prime}(\epsilon_{{\bf k}}) \biggr[\frac{\partial^2 \epsilon_{\bf k}}{\partial k_x^2} \frac{\partial^2 \epsilon_{\bf k}}{\partial k_y^2}  - \biggr(\frac{\partial^2 \epsilon_{\bf k}}{\partial k_x \partial k_y}\biggr)^2 \biggr], \\
&=& -\frac{2e^2a^4}{3\hbar^2 c^2}t \sum_{{\bf k}}\delta((\epsilon_k-\mu)/t_0)\cos{k_xa}\cos{k_ya},  \label{LP} \\
&=& -\frac{2e^2a^4}{3\pi^2 \hbar^2 c^2}t(E(\kappa)-\frac{1}{2}K(\kappa)), 
\end{eqnarray}
where $\kappa=\sqrt{1-\mu^2/16t^2}$, $\epsilon_k= -2t_0(\cos{(k_xa)}+\cos{(k_ya)})$, and $\mu$ is the chemical potential and $K(k)$ and $E(k)$ are elliptic integrals of the first and second kind, respectively~\cite{Ogata2}.

On the other hand, the $t_2$ term is already proportional to $h^2$.
Therefore, we can neglect the effect of the Peierls phase, $\Phi_{ij}$, for this term when we calculate the ground-state energy up to the second order with respect to the magnetic field.
As a result, the correction of the ground-state energy is given by
\begin{eqnarray}
\Delta E_2 =  4t_2h^2\sum_{{\bf k}}\bigr[ \cos{(k_xa)} +\cos{(k_ya)} \bigr],  \label{c0}
\end{eqnarray}
where $\sum_{{\bf k}}$ is the summation of the occupied states.
This correction gives a contribution to the susceptibility of
\begin{eqnarray}
\chi_2 &=& -\frac{2e^2a^4}{c^2\hbar^2}t_2\sum_{{\bf k}}\bigr[ \cos{(k_xa)} +\cos{(k_ya)} \bigr], \label{correction0} \\
&=& -\frac{4e^2a^4}{\pi^2c^2\hbar^2}t_2\bigr[ E(\kappa) -(1-\kappa^2)K(\kappa)\bigr]. \label{correction}
\end{eqnarray}
Since $\chi_2$ originates from the correction of the hopping integral due to the magnetic field, this correction term of the Peierls phase is the Fermi sea contribution.
Note that the sign of $\chi_2$ is dependent on those of $t$ and $t_2$.
When $t$ and $t_2$ have the same (a different) sign, the hopping $t(h)$ decreases (increases) and the absolute value of the total energy decreases (increases).
As a result, diamagnetic (paramagnetic) susceptibility occurs.

\begin{figure}
\rotatebox{0}{\includegraphics[angle=0,width=1\linewidth]{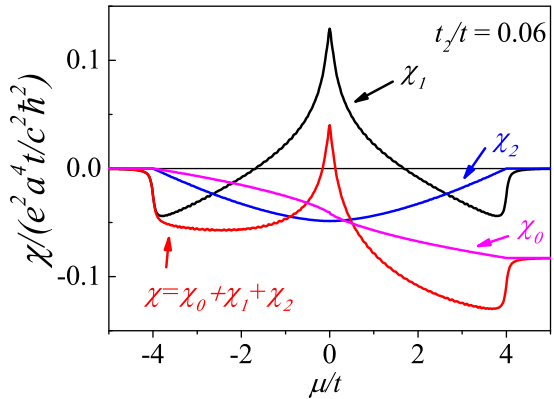}}
\caption{(Color online) Chemical potential dependence of orbital susceptibility on square lattice for $t_2/t= 0.06$.
$\chi_0$, $\chi_1$, $\chi_2$, and $\chi$ are the orbital susceptibilities of the site diagonal term, the Landau-Peierls term, and the correction of the Peierls phase, and the total orbital susceptibility ($\chi = \chi_0 +\chi_1 + \chi_2$), respectively.
}
\label{square}
\end{figure}

Figure \ref{square} shows the chemical potential dependence of the orbital susceptibility on the square lattice for $t_2/t_0= 0.06$.
The pink, black, and blue lines indicate the orbital susceptibilities given by eqs.\ (\ref{site_dia}), (\ref{LP}), and (\ref{correction0}), respectively.
The red line indicates the total orbital susceptibility $\chi = \chi_0+ \chi_1 +\chi_2$, where, to estimate $\chi_0$, we use $a=2.684a_B$, $Z^*=3.25$, and $t =3.6$ eV as discussed in Sect.\ 3.
It is found that the diamagnetic region is expanded by the correction of the Peierls phase ($\chi_2$).

Finally, we comment on the present results for the square lattice compared with those obtained from the general formula in terms of Bloch wave functions~\cite{Ogata,Ogata2}.
Although the 1s orbital case was studied in detail~\cite{Ogata2}, it is straightforward to apply the same method to the 2p$_{\pi}$ orbital.
As a result, the result in Fig. \ref{square} is consistent with the recent paper~\cite{Ogata2}.

\section{Conclusion}
We have developed an extended formula for orbital susceptibility including corrections of the Peierls phase by extending Pople's method. 
As a first step, we estimated the orbital susceptibility of benzene on the basis of the $\pi$-electron approximation.
As a result, we analytically showed that the orbital susceptibility is 1.2 times larger than that estimated only from the Peierls phase.
Next, we calculated the Coulomb interaction dependence of the orbital susceptibility of benzene by exact diagonalization.   
We found that as the Coulomb interaction increases, the absolute value of the orbital susceptibility decreases, while the ratio of the orbital susceptibility with and without the correction of the Peierls phase increases.
We expect that the orbital susceptibility will become closer to the experimental result when we consider the correction. 
Finally, we calculated the orbital susceptibility of a single-band tight-binding model on a square lattice.
We showed that the corrections of the Peierls phase give a contribution to orbital susceptibility comparable to the Landau-Peierls contribution.
We also clarified that the correction of the Peierls phase corresponds to Fermi sea term in the tight-binding model.   
The obtained result is in very good agreement with the previous result obtained from the exact formula based on general Bloch bands~\cite{Ogata,Ogata2}.

\begin{acknowledgments}
One of the authors (H.M.) thanks I. Proskurin and H. Fukuyama for the discussions.
This work was supported by the JSPS Core-to-Core Program, A. Advanced Research Networks, and a Grant-in-Aid for Scientific Research on Innovative AreasgUltra Slow Muon Microscopeh(No. 23108004) from the Ministry of Education, Culture, Sports, Science and Technology, Japan.
We were also supported by Grants-in-Aid for Scientific Research from the Japan Society for the Promotion of Science (No. 15K17694, No. 25220803, and No. 15H02108).
 \end{acknowledgments}


\def\journal#1#2#3#4{#1 {\bf #2}, #3 (#4) }
\def\PR{Phys.\ Rev.}
\def\PRB{Phys.\ Rev.\ B}
\def\PRL{Phys.\ Rev.\ Lett.}
\def\JPSJ{J.\ Phys.\ Soc.\ Jpn.}
\def\PTP{Prog.\ Theor.\ Phys.}
\def\JPCS{J.\ Phys.\ Chem.\ Solids}

\end{document}